\documentclass[useAMS,usenatbib]{mn2e}

\usepackage{epsfig}
\usepackage{textcomp}
\usepackage{amssymb,amsmath}
\usepackage{longtable}
\usepackage{lscape}

\newcommand{\xmm} {{\it XMM-Newton}}
\newcommand{\chandra} {{\it Chandra}}

\newcommand{\cmsq} {cm$^{-2}$}
\newcommand{\nh} {$N_{\rm{H}}$}
\newcommand{\lx} {$L_{\rm{X}}$}

\newcommand{\mic}{{${\umu}$m}}

\newcommand{\mgii}{{\rm{Mg\,\sc{ii}}}}
\newcommand{\ha}{{\rm{H$\alpha$}}}
\newcommand{\hb}{{\rm{H$\beta$}}}

\newcommand{\ergs}{\mbox{\thinspace erg\thinspace s$^{-1}$}}
\newcommand{\rs} {$r_{\rm S}$}
\newcommand{\lbol} {$L_{\rm Bol}$}
\newcommand{\mbh} {$M_{\rm BH}$}
\newcommand{\aox} {$\alpha_{\rm OX}$}
\newcommand{\lamedd} {$\lambda_{\rm Edd}$}

\newcommand{\redrangea}{$0.5<z<1.2$}
\newcommand{\redrangeb}{$1.2<z<2.1$}
\newcommand{\nbhxdata}{260}
\newcommand{\nbhbxdata}{79}
\newcommand{\nbhbxrqdata}{73}

\newcommand{\neddsedxdata}{44}
\newcommand{\neddaoxxdata}{31}
\newcommand{\neddxrqdata}{69}

\newcommand{\nlbolsed}{ 207}
\newcommand{\reggamlam}{$\Gamma=(0.32\pm0.05)$log$_{10}$\lamedd$+(2.27\pm0.06)$}
\newcommand{\reggamlamm}{$0.32\pm0.05$}
\newcommand{\reggamlamc}{$2.27\pm0.06$}
\newcommand{\reggamlamha}{$\Gamma=(0.34\pm0.07)$log$_{10}$\lamedd$+(2.34\pm0.09)$}
\newcommand{\reggamlamham}{$0.34\pm0.07$}

\newcommand{\reggamlammg}{$\Gamma=(0.35\pm0.05)$log$_{10}$\lamedd$+(2.28\pm0.06)$}
\newcommand{\reggamlammgm}{$0.35\pm0.05$}
\newcommand{\reggamlammgc}{$2.28\pm0.06$}
\newcommand{\pnhgamlam}{6.59$\times10^{-8}$}
\newcommand{\rsgamlam}{0.60}

\newcommand{\reggamfwhm}{$\Gamma=( -0.69\pm  0.11)$log$_{10}$(FWHM/km s$^{-1}$)$+(  4.44\pm  0.42)$}
\newcommand{\reggamfwhmha}{$\Gamma=( -0.49\pm  0.14)$log$_{10}$(FWHM/km s$^{-1}$)$+(  3.72\pm  0.53)$}
\newcommand{\reggamfwhmmg}{$\Gamma=( -1.02\pm  0.15)$log$_{10}$(FWHM/km s$^{-1}$)$+(  5.70\pm  0.54)$}
\newcommand{\pnhgamfwhm}{2.71$\times10^{-4}$}
\newcommand{\pnhgamfwhmmg}{1.32$\times10^{-4}$}
\newcommand{\rsgamfwhm}{ -0.41}

\newcommand{\nkbolsample}{          32}

\newcommand{\radiodet}{          21}

\title[]{A statistical relation between the X-ray spectral index and Eddington ratio of active galactic nuclei in deep surveys}

\author[M. Brightman, et al.]{M. Brightman$^{1}\thanks{E-mail: mbright@mpe.mpg.de}$, J. D. Silverman$^{2}$, V. Mainieri$^{3}$, Y. Ueda$^{4}$, M. Schramm$^{2}$, K. Matsuoka$^{5,6}$ \and 
T. Nagao$^{7,4}$, C. Steinhardt$^{2}$, J. Kartaltepe$^{8}$, D. B. Sanders$^{9}$, E. Treister$^{10}$, O. Shemmer$^{11}$ \and 
W. N. Brandt$^{12}$, M. Brusa$^{1,13,14}$,  A. Comastri$^{14}$, L. C. Ho$^{15}$, G. Lanzuisi$^{1}$, E. Lusso$^{16}$ \and
K. Nandra$^{1}$, M. Salvato$^{1}$, G. Zamorani$^{14}$, M. Akiyama$^{17}$, D. M. Alexander$^{18}$, A. Bongiorno$^{1,19}$ \and  
P. Capak$^{20}$, F. Civano$^{21}$, A. Del Moro$^{18}$, A. Doi$^{22}$, M. Elvis$^{21}$, G. Hasinger$^{9}$, E. S. Laird$^{23}$  \and  
D. Masters$^{15}$,  M. Mignoli$^{14}$, K. Ohta$^{4}$, K. Schawinski$^{24}$, Y. Taniguchi$^{6}$\\
$^{1}$Max-Planck-Institut f\"{u}r extraterrestrische Physik, Giessenbachstrasse 1, D-85748, Garching bei M\"{u}nchen, Germany\\
$^{2}$Kavli Institute for the Physics and Mathematics of the Universe (WPI), Todai Institutes for Advanced Study, the University of Tokyo\\
$^{3}$European Southern Observatory, Karl-Schwarzschild-Strasse 2, 85748 Garching bei M\"{u}nchen, Germany\\
$^{4}$Department of Astronomy, Kyoto University, Kitashirakawa-Oiwake- cho, Sakyo-ku, Kyoto 606-8502, Japan\\
$^{5}$Department of Physics and Astronomy, Seoul National University, 599 Gwanak-ro, Gwanak-gu, Seoul 151-742, Republic of Korea\\
$^{6}$Research Center for Space and Cosmic Evolution, Ehime University, 2-5 Bunkyo-cho, Matsuyama 790-8577, Japan\\
$^{7}$The Hakubi Center for Advanced Research, Kyoto University, Yoshida- Ushinomiya-cho, Sakyo-ku, Kyoto 606-8302, Japan\\
$^{8}$National Optical Astronomy Observatory, 950 North Cherry Aveue, Tucson, AZ 85719, USA\\
$^{9}$Institute for Astronomy, University of Hawaii, 2680 Woodlawn Drive, Honolulu, HI 96822, USA\\
$^{10}$Departamento de Astronom\'{i}a, Universidad de Concepci\'{o}n, Casilla 160- C, Concepci\'{o}n, Chile\\
$^{11}$Department of Physics, University of North Texas, Denton, TX 76203, USA\\
$^{12}$Department of Astronomy and Astrophysics, The Pennsylvania State University, University Park, PA 16802, USA\\
$^{13}$Dipartimento di Fisica e Astronomia, Universit\'{a} degli Studi di Bologna, viale Berti Pichat 6/2, 40127 Bologna, Italy\\
$^{14}$INAF Ð Osservatorio Astronomico di Bologna, Via Ranzani 1, 40127 Bologna, Italy\\
$^{15}$The Observatories of the Carnegie Institution for Science, 813 Santa Barbara Street, Pasadena, CA 91101, USA\\
$^{16}$Max Planck Institut f\"{u}r Astronomie, K\"{o}nigstuhl 17 D-69117 Heidelberg, Germany\\
$^{17}$Astronomical Institute, Tohoku University, 6-3 Aramaki, Aoba-ku, Sendai 980-8578, Japan\\
$^{18}$Department of Physics, Durham University, South Road, Durham DH1 3LE, UK\\
$^{19}$INAF Ð Osservatorio Astronomico di Roma, Via di Frascati 33, 00040 Monte Porzio Catone, Italy\\
$^{20}$NASA/JPL Spitzer Science center, California Institute of Technology, 1200 East California Boulevard, Pasadena, CA 91125, USA\\
$^{21}$Harvard-Smithsonian Center for Astrophysics, 60 Garden Street, Cambridge, MA 02138, USA\\
$^{22}$The Institute of Space and Astronautical Science, Japan Aerospace Exploration Agency, 3-1-1 Yoshinodai, Chuou-ku, Sagamihara, Kanagawa 252-5210, Japan\\
$^{23}$Astrophysics Group, Imperial College London, Blackett Laboratory, Prince Consort Road, London SW7 2AZ, UK\\
$^{24}$Department of Physics, Yale University, New Haven, CT 06520, USA}

\begin{document}

\maketitle

\label{firstpage}

\clearpage

\begin{abstract}
We present an investigation into how well the properties of the accretion flow onto a supermassive black hole may be coupled to those of the overlying hot corona.  To do so, we specifically measure the characteristic spectral index, $\Gamma$, of a power-law energy distribution, over an energy range of 2 to 10 keV, for X-ray selected, broad-lined radio-quiet active galactic nuclei (AGN) up to z$\sim$2 in COSMOS and E-CDF-S.  We test the previously reported dependence between $\Gamma$ and black hole mass, FWHM and Eddington ratio using a sample of AGN covering a broad range in these parameters based on both the \mgii\ and \ha\ emission lines with the later afforded by recent near infrared spectroscopic observations using Subaru/FMOS. We calculate the Eddington ratios, \lamedd, for sources where a bolometric luminosity (\lbol) has been presented in the literature, based on SED fitting, or, for sources where these data do not exist, we calculate \lbol\ using a bolometric correction to the X-ray luminosity, derived from a relationship between the bolometric correction, and \lx/$L_{3000}$. From a sample of \neddxrqdata\ X-ray bright sources ($>$ 250 counts), where $\Gamma$ can be measured with greatest precision, with an estimate of \lbol, we find a statistically significant correlation between $\Gamma$ and \lamedd, which is highly significant with a chance probability of \pnhgamlam. A statistically significant correlation between $\Gamma$ and the FWHM of the optical lines is confirmed, but at lower significance than with \lamedd\ indicating that \lamedd\ is the key parameter driving conditions in the corona. Linear regression analysis reveals that \reggamlam\ and  \reggamfwhm. Our results on $\Gamma$-\lamedd\ are in very good agreement with previous results. While the $\Gamma$-\lamedd\ relationship means that X-ray spectroscopy may be used to estimate black hole accretion rate, considerable dispersion in the correlation does not make this viable for single sources, however could be valuable however for large X-ray spectral samples, such as those to be produced by {\it eROSITA}. 

\end{abstract}

\section{Introduction}
X-ray emission from AGN is ubiquitous \citep{tananbaum79}, and is often used itself as an indicator of black hole accretion activity. The hard X-ray ($>2$ keV) spectrum takes the form of, at least to first order, a power-law, where the photon flux, $F_{\gamma}=AE^{-\Gamma}$ (photons cm$^{-2}$ s$^{-1}$). The X-ray emission is thought to be produced by the Compton up-scattering of seed optical/UV photons, produced by thermal emission from the accretion disc \citep{shakura73}. This is believed to be done by hot electrons forming a corona in proximity to the disc \citep[e.g.][]{sunyaev80}. Investigations into the X-ray emission can yield important insights into the accretion process and constrain accretion models \citep[e.g.][]{haardt91,haardt93}. However, many details regarding this process remain unclear, for example the geometry of the corona, its heating and the energy transfer between the two phases. 

The discovery of the intrinsic power-law index measured to be $\Gamma\sim1.9$ was an important step in supporting the disc-corona model \citep{pounds90,nandra94}, as previous results presenting lower values \citep[e.g.][]{mushotsky84,turner89} were a challenge for these models. Subsequent studies have tried to pin down the parameters of the accretion which give rise to the physical conditions of the corona, such as the electron temperature, T$_{e}$ and optical depth to electron scattering, $\tau_{\rm es}$, upon which $\Gamma$ depends \citep{rybicki86}. 

The temperature, and thus emission spectrum of a standard Shakura \& Sunyaev accretion disc depends on the mass accretion rate, $\dot{m}$, \citep{shakura73}. The luminosity of the system is thus related to $\dot{m}$ via the accretion efficiency, $L_{\rm Bol}=\eta\dot{m}c^{2}$, which is often parametrised as a fraction of the Eddington luminosity, $L_{\rm Edd}$, by the Eddington ratio, \lamedd=$L_{\rm Bol}/L_{\rm Edd}$. $L_{\rm Edd}$ is the theoretical maximal luminosity achieved via accretion when accounting for radiation pressure and is dependent on the black hole mass ($L_{\rm Edd}=4\pi$ G\mbh\ $m_{\rm p}c/\sigma_{\rm T}\simeq1.26\times10^{38}$\mbh\ \ergs, where \mbh\ is the black hole mass in solar masses). Due to the dependance of the accretion disc spectrum on these parameters, several works have investigated how the coronal X-ray emission is coupled. It has been shown that $\Gamma$ is well correlated with the full width at half maximum (FWHM) of the broad optical emission lines, specifically \hb\ \citep[e.g.][]{boller96,brandt97}, giving some indication that the gravitational potential, i.e. black hole mass, is key. Subsequently \lamedd\ was shown to be strongly correlated with $\Gamma$ \citep{lu99,wang04,shemmer06}. However a known degeneracy exists between the \hb\ FWHM and \lamedd\ of the system \citep{boroson92}, making it difficult to determine the fundamental parameter behind these relationships. \cite{shemmer06}(S06) were able to break this degeneracy by adding highly luminous quasars to their analysis, concluding that \lamedd\ is the primary parameter driving the conditions in the corona, giving rise to $\Gamma$. \cite{shemmer08} (S08) followed up the work of S06 by adding further luminous quasars, increasing the significance of the previous result.

 Follow up studies have confirmed this relationship using larger samples and extending the range of parameters probed \citep[e.g.][]{risaliti09b,jin12,fanali13}. Furthermore, works into the possible dependence of $\Gamma$ on X-ray luminosity (\lx) have reported a positive correlation between these quantities in high redshift sources \citep{dai04,saez08}, but not seen in the local universe \citep{george00,brightman11}, and an evolution of this correlation was also reported in \cite{saez08}. One interpretation of this correlation, however, was that it is driven fundamentally by changes in $\dot{m}$. It has been shown that $\dot{m}$ can also be estimated from X-ray variability analysis \citep[e.g.][]{mchardy06}, where $\dot{m}$ is correlated with the break timescale in the power density spectrum. \cite{papadakis09} used this result when conducting a joint X-ray spectral and timing analysis of nearby Seyfert galaxies to show in an independent manner that $\Gamma$ correlates with $\dot{m}$. Furthermore, in a study of the X-ray variability in primarily local AGN, \cite{ponti12} find a significant correlation between the excess variance, $\sigma^2_{\rm rms}$, and $\Gamma$, which when considering the correlation between $\sigma^2_{\rm rms}$ and $\dot{m}$ that they find, is indirect evidence for the dependance of $\Gamma$ on $\dot{m}$.

The strong correlation between $\Gamma$ and accretion rate is interpreted as enhanced emission from the accretion disc in higher accretion rate systems more effectively cooling the corona, leading to a steepening of the X-ray emission \citep{pounds95}. 

Not only is the $\Gamma$-\lamedd\ correlation significant with respect to constraining accretion models, but as pointed out by S06, it allows an independent measurement of the black hole growth activity in galaxies from X-ray spectroscopy alone, and with a measurement of the bolometric luminosity, a black hole mass can be determined. This would be especially useful for moderately obscured AGN, where virial black hole mass estimates are not possible, but X-rays can penetrate the obscuration. 

The aim of this work is to extend previous analyses on correlations with $\Gamma$ for the first time to the deep extragalactic surveys and in doing so, extending the range of parameters explored, breaking degeneracies between parameters where possible. By the use of survey data, we benefit from uniform X-ray data, where previous studies have relied on non-uniform archival data. In order to use $\Gamma$ as an Eddington ratio indicator, the dispersion in the relationship must be well parametrised, which we aim to do with the large range in parameters explored and our uniform X-ray coverage.

In addition, our black hole mass estimates are based on two optical lines, \ha\ and \mgii. \cite{risaliti09b} (R09) showed that there was a stronger correlation with \lamedd\ for measurements made with \hb\ compared to \mgii\ and C {\sc iv}, suggesting that this line is the best black hole mass indicator of the three. The \ha\ data used here were specifically pursued partly in order to investigate the $\Gamma$ correlations with \ha\ data for the first time at high redshifts, facilitated by near-infrared (NIR) spectroscopy. The parameters we explore in this work are $L_{\rm UV}$, \lx, FWHM, \mbh\ and \lamedd.

In this work we assume a flat cosmological model with $H_{\rm 0}$=70 km s$^{-1}$ Mpc$^{-1}$ and $\Omega_{\Lambda}$=0.70. For measurement uncertainties on our spectral fit parameters we present the 90\% confidence limits given two interesting parameters ($\Delta$c-stat=4.61).

\section{Sample properties and data analysis}

\subsection{Sample selection}

\begin{figure}
\includegraphics[width=90mm]{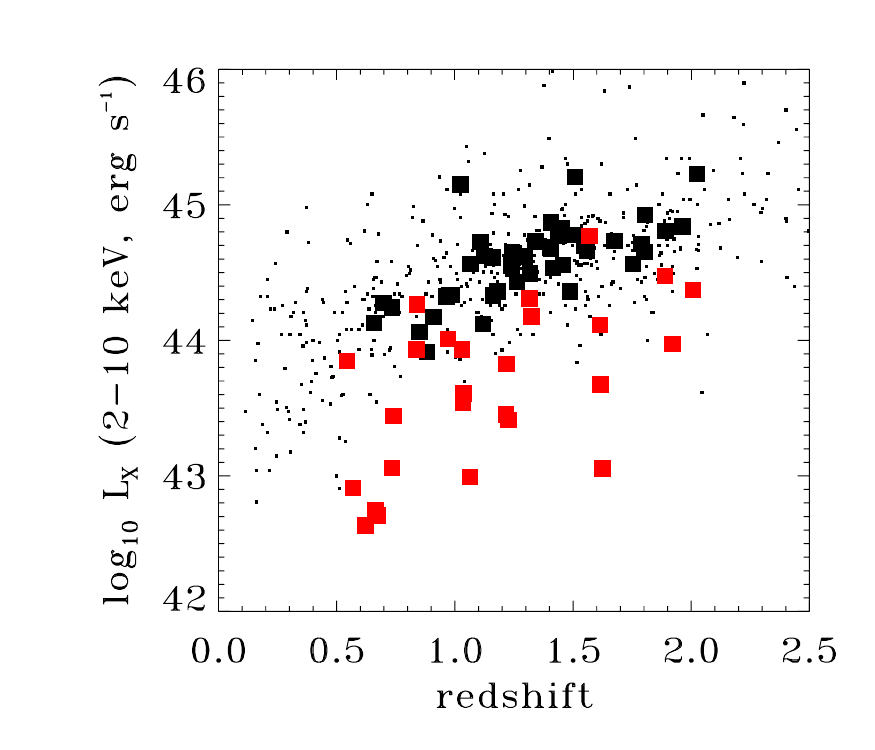}
\caption{Plot of the redshift and X-ray luminosity distribution in our combined sample, where large red data points are from the E-CDF-S  and the large black data points are from COSMOS. The sample covers redshifts up to $\sim2.1$ and 2-10 keV X-ray luminosities of $42.5\lesssim$ log$L_{\rm X}$/erg s$^{-1}\lesssim45.5$. The dots show the sample of R09, which is derived from the {\it SDSS/XMM-Newton} quasar survey of Young, et al (2009).}
\label{zlx}
\end{figure}

Our sample is based on two major extragalactic surveys, the extended {\it Chandra} Deep Field-South \citep[E-CDF-S][]{lehmer05}, inclusive of the ultra-deep central area \citep{giacconi02,luo08,xue11}, and COSMOS surveys \citep{cappelluti09,elvis09}, where deep X-ray coverage with high optical spectroscopic completeness exists in both fields. The sample selection is as follows:
\begin{itemize}
\item We select sources with a black hole mass estimates based on \ha\ or \mgii\ from a black hole mass catalogue to be presented in Silverman, et al (in preparation).  Mg {\sc ii} line measurements were made from extensive existing optical spectra, primarily from zCOSMOS, Magellan/IMACS, Keck/Deimos and SDSS. Measurements of \ha\ up to $z\sim1.8$ are facilitated by the use of NIR spectroscopy from the Fiber Multi Object Spectrograph (FMOS) on the Subaru Telescope. The targets of these observations were {\it Chandra} X-ray selected AGN in COSMOS \citep{elvis09} and E-CDF-S \citep{lehmer05} with an already known spectroscopic redshift and detection of a broad emission line (FWHM$>$2000 km s$^{-1}$). A subset of the optical and NIR spectroscopic data analysis has already been presented in \cite{matsuoka13}. 
\item In COSMOS we select Chandra sources also detected by XMM-Newton \citep[see discussion in][]{brusa10}, due to the higher throughput capability of this observatory with respect to Chandra. For the E-CDF-S, we use the deep Chandra data available.
\end{itemize}
Combining these surveys and the selection criteria above yields a sample of \nbhxdata\ sources with both a black hole mass estimate and X-ray data, however, the main goal of this work is to investigate the detailed relationship between the coronal X-ray emission, characterised by the power-law index, $\Gamma$, and the parameters of the accretion, the observed quantities being luminosity and FWHM, and the derived quantities being \mbh\ and \lamedd. We therefore make the following cuts to the above sample as follows:
\begin{itemize}
\item We take sources where there are at least 250 source counts in the X-ray spectrum in order to get an accurate measurement of $\Gamma$. We describe this further in section \ref{xspec}. 
\item For analysis with \lamedd, a bolometric luminosity is required which we take from \cite{lusso12} (L12), which are derived from spectral energy distribution (SED) fitting. As these data do not exist for E-CDF-S  sources, we use a bolometric correction to the X-ray luminosity, derived from \lx/$L_{3000}$, which we describe in section \ref{lbolest}. There are \neddsedxdata\ which have an \lbol\ from L12, all of which are in COSMOS, and \neddaoxxdata\ of which have \lbol\ calculated using \lx/$L_{3000}$, all of which are in the E-CDF-S.
\item As we wish to study the properties of the coronal emission, responsible for the majority of X-ray emission in radio quiet AGN, we exclude radio loud sources. For this we make a cut of R$<$100, where R is the radio loudness parameter \citep{kellerman89}, excluding six sources. We describe the radio properties of our sample in section \ref{radio}. Our final sample consists of \nbhbxrqdata\ sources for analysis with FWHM and \mbh\ and \neddxrqdata\ sources for analysis with \lamedd. 

Fig \ref{zlx} shows how our final sample spans the redshift-luminosity plane, and how it compares to the sample of R09 which is derived from the {\it SDSS/XMM-Newton} quasar survey of \cite{young09}. Our combination of wide and deep survey data allows us to span a larger luminosity range than done previously, and better sample to luminosity-redshift plane.
\end{itemize}

\subsection{Emission line measurements and black hole mass estimation}

The first step in our analysis is to determine the accretion parameters, the velocity dispersion of the material around the black hole, characterised by the FWHM of the optical lines and \mbh, which we derive from single-epoch optical spectral line fitting. A full description of the target selection, data analysis, including spectral line fitting, and results on virial black hole mass is presented in \cite{matsuoka13}, which we briefly describe here.

Mg {\sc ii} line measurements were made from optical spectra obtained primarily from zCOSMOS, Magellan/IMACS and SDSS. Fitting of the \mgii\ line was carried out using a power-law to characterise the continuum, and one to two Gaussians used for the line. A broad Fe emission component is also included, which is based on the empirical template of \cite{vestergaard01}.

The \ha\ measurements used here were the result of a campaign to make Balmer line observations of AGN outside the optical window at $z>0.8$ in the NIR with Subaru/FMOS.  FMOS observations were made in low resolution mode ($\Delta\lambda/\lambda\simeq600$) simultaneously in the J (1.05-1.34 \mic) and H (1.43-1.77 \mic) bands, yielding a velocity resolution of FWHM$\sim$500 km s$^{-1}$ at $\lambda=1.5$ \mic. The target selection was such that the continuum around \ha\ could be accurately determined and fit by a power-law. The \ha\ line was fit by 2 to 3 Gaussians, while the neighbouring [NII]$\lambda$6548,6684 lines were fit with a pair of Gaussians.  

Virial black hole estimates were calculated with the following formula:

\begin{equation}
{\rm log}\frac{M_{\rm BH}}{M_{\odot}}=a+b\,{\rm log}\frac{\lambda L_{\lambda}\, \rm{or}\,  L_{\rm line}}{10^{44} \ergs}+c\, \rm{log}\frac{\rm FWHM}{\rm km s^{-1}}
\end{equation}
where a=1.221, b=0.550 and c=2.060 for \ha\ from \cite{greene05} and a=0.505, b=0.620 and c=2.000 for \mgii\ from \cite{mclure02}. For \ha, the line luminosity is used in the calculation, whereas for \mgii\ the monochromatic luminosity at 3000 \AA\ is used.

In \cite{matsuoka13} a comparison of virial black hole mass estimates from \ha\ and \mgii\ is presented. They find a tight correlation between $L_{\rm H\alpha}$ and $L_{3000}$ and a close one-to-one relationship between the FWHM of \ha\ and \mgii, therefore leading to good agreements between the mass estimates from these lines. While \hb\ detections with FMOS do exist for this sample, measurements with this line is to be presented in a forthcoming publication (Silverman, et al. in preparation).

\subsection{X-ray Data: Chandra Deep Field-South}
\label{cdfs}
In the E-CDF-S, the targets of the FMOS observations were the optical counterparts of type 1 AGN detected in the E-CDF-S survey described by \cite{lehmer05}. The E-CDF-S consists of nine individual observations with four different central pointing positions, to a depth of 250 ks. The central region of the E-CDF-S survey is the location of the ultra deep 4 Ms E-CDF-S  survey, which is described by \cite{xue11} and consists of 52 individual observations with a single central pointing position. We utilise all the \chandra\ data in this region. The data were screened for hot pixels and cosmic afterglows as described in \cite{laird09}, astrometric corrections made as described in \cite{rangel13} and the source spectra were extracted using the {\sc acis extract} (AE) software package \footnote{The {\sc acis extract} software package and User's Guide are available at http://www.astro.psu.edu/xray/acis/acis\_analysis.html} \citep{broos10}, using the positions of the E-CDF-S sources. AE extracts spectral information for each source from each individual observation ID (obsID) based on the shape of the local point spread function (PSF) for that particular position on the detector. We choose to use regions where 90\% of the PSF has been enclosed at 1.5 keV. Background spectra are extracted from an events list which has been masked of all detected point sources in \cite{xue11} and \cite{lehmer05}, using regions which contain at least 100 counts. AE also constructs response matrix files (RMF) and auxiliary matrix files (ARF). The data from each obsID are then merged to create a single source spectrum, background spectrum, RMF and ARF for each source. The rest-frame 2-10 keV signal-to-noise ratios in this field range from 2.5 to 500.

\subsection{X-ray Data: XMM-COSMOS}

In the COSMOS survey, the targets of the FMOS observations were the optical counterparts of type 1 AGN detected in the \chandra-COSMOS survey \citep{elvis09,civano12}. However, as medium-depth ($\sim$60ks) \xmm\ data exist in this field, we take advantage of the high throughput of this satellite to obtain X-ray spectral data. The XMM-COSMOS data are described in \cite{cappelluti09} and the procedure adopted to extract sources and background spectra in \cite{mainieri07}. We briefly recall the main steps here. The task {\it region} of the \xmm\ Science Analysis System (SAS)\footnote{http://xmm.vilspa.esa.es/external/xmm\_sw\_cal/sas\_frame.shtml} software has been used to generate the source and background extraction regions. The source region is defined as a circle with radius r$_s$ that varies according to the signal-to-noise and the off-axis angle of the detection to optimise the quality of the final spectrum. The radii of these regions are reduced by the task to avoid overlapping with the extraction regions of nearby sources. All source regions are further excised from the area used for the background measurement. The task {\it especget} has been used to extract from the event file the source and background spectra for each object. The same task generates the calibration matrices (i.e. arf and rmf) for each spectrum and determines the size of the source and background areas while updating the keyword BACKSCAL in the header of the spectra appropriately\footnote{The header keyword   BACKSCAL is set to 1 for the source spectrum while for the background spectrum it is fixed to the ratio between the background to source areas.}. The single pointing spectra have been combined with {\it mathpha} to generate the spectrum of the whole observation.\footnote{We note that all the XMM-Newton observations in the COSMOS field have been performed with the thin filter for the {\it pn} camera.} The rest-frame 2-10 keV signal-to-noise ratios in this field range from 2.5 to 16.

\subsection{X-ray spectral analysis}
\label{xspec}

\begin{figure}
\includegraphics[width=90mm]{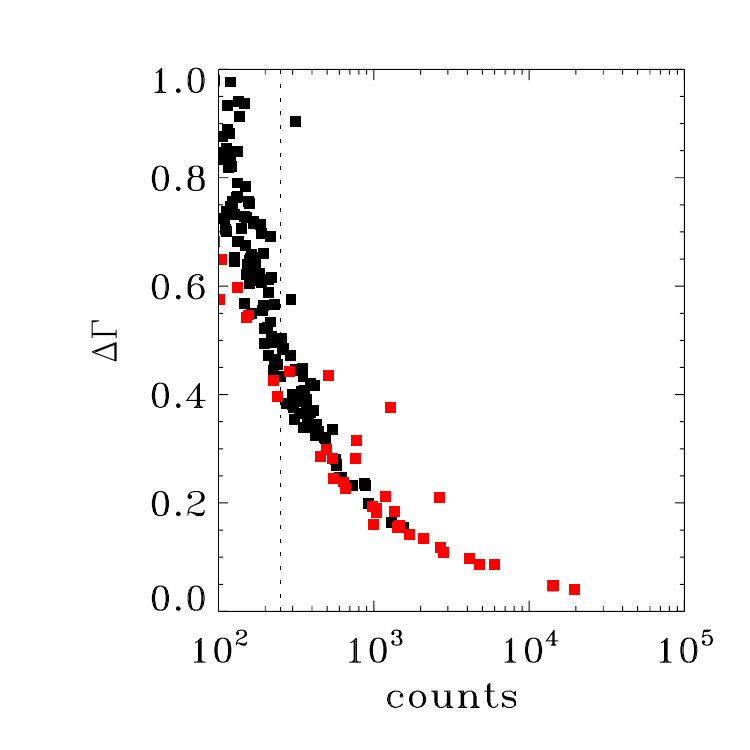}
\caption{Plot of the uncertainty of $\Gamma$ as a function of the number of source counts in the X-ray spectrum. We cut the sample to sources with greater than 250 counts in order to include sources for which $\Delta\Gamma<0.5$. Red data points show where the data come from the E-CDF-S , while black data points come from COSMOS. The points lying above the main relation are those where absorption has been included in the fit.}
\label{gamma2}
\end{figure}

The goal of the X-ray spectral analysis is to uniformly measure $\Gamma$ and \lx\ in the rest-frame 2-10 keV range for all \nbhxdata\ sources with black hole mass estimates. For both \chandra\ and \xmm\ data, we lightly bin the spectral data with at least one count per bin using the {\sc heasarc} tool {\tt grppha}. We use {\sc xspec} version 12.6.0q to carry out X-ray spectral fitting, and the Cash statistic \citep[cstat,][]{cash79} as the fit statistic. We fit the rest-frame 2-10 keV spectra with a power-law model, neglecting the rest-frame 5.5-7.5 keV data where the iron K complex is emitted. We use all \nbhxdata\ sources for analysis with \lx, however, measurement of $\Gamma$ requires higher quality spectral data in order to obtain good constraints. Fig. \ref{gamma2} shows how the uncertainty on $\Gamma$ decreases as the number of counts in the spectrum increases. We restrict analysis with $\Gamma$ to \nbhbxdata\ sources where there are greater than 250 counts in the X-ray spectrum, which are needed in order to measure $\Gamma$ to $\Delta\Gamma<0.5$. In the spectral fit, we include a local absorption component, {\tt wabs}, to account for Galactic absorption, with \nh=$7\times10^{19}$ \cmsq\ for the E-CDF-S  and \nh=$1.7\times10^{20}$ \cmsq\ for the COSMOS. By restricting our analysis to rest frame energies above 2 keV, we are insensitive to \nh\ values below$\sim10^{22}$ \cmsq. As we are studying broad-lined type 1 AGN, absorption intrinsic to the source is not generally expected above this, however, we assess this by adding an absorption component in the spectral fit ({\tt zwabs} in {\sc xspec}), and noting any improvement in the fit statistic. Based on an F-test, we find evidence at $>$90\% confidence for absorption in three sources, for which we use the measurements carried out with the addition of the absorption component. These are E-CDF-S IDs 367 (\nh$=1.6\times10^{22}$ \cmsq) and 391 (\nh$=6.9\times10^{21}$ \cmsq) and XMM-COSMOS ID 57 (\nh$=2.9\times10^{22}$ \cmsq). For the remaining sources, spectral fits were carried out with a simple power-law component. Recent work by \cite{lanzuisi13} (L13) have presented spectral analysis of bright X-ray sources ($>$70 counts) in \chandra-COSMOS, however utilising the full 0.5-7 keV \chandra\ bandpass. We compare our results to theirs in a later section.

\subsection{Determining \lamedd}
\label{lbolest}

\begin{figure*}
\includegraphics[width=90mm]{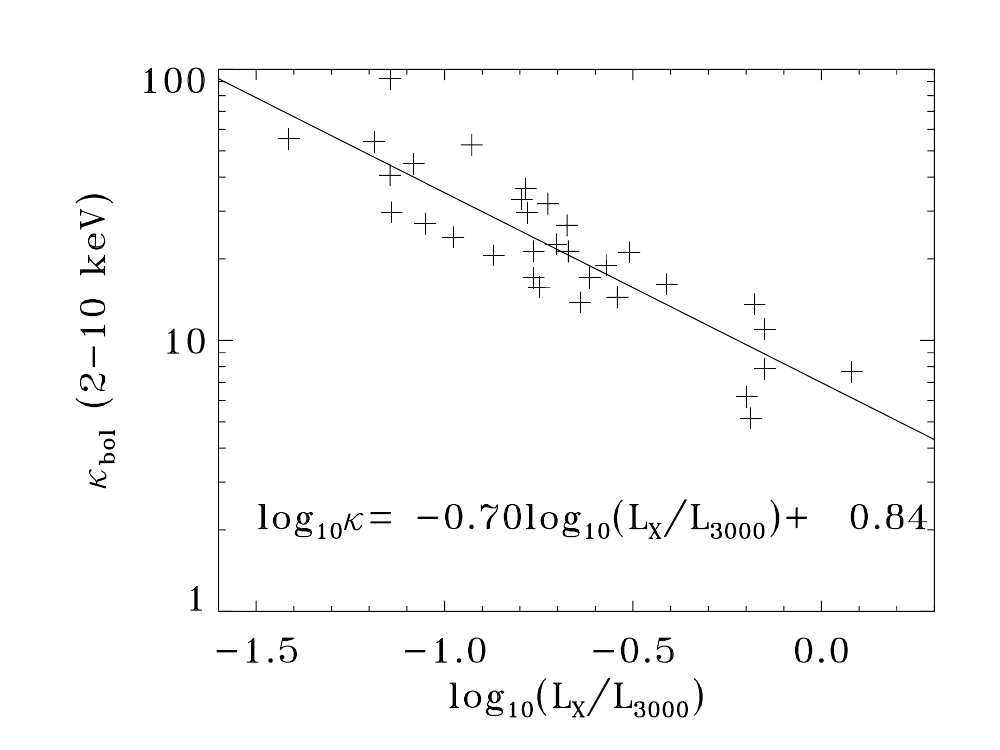}\includegraphics[width=90mm]{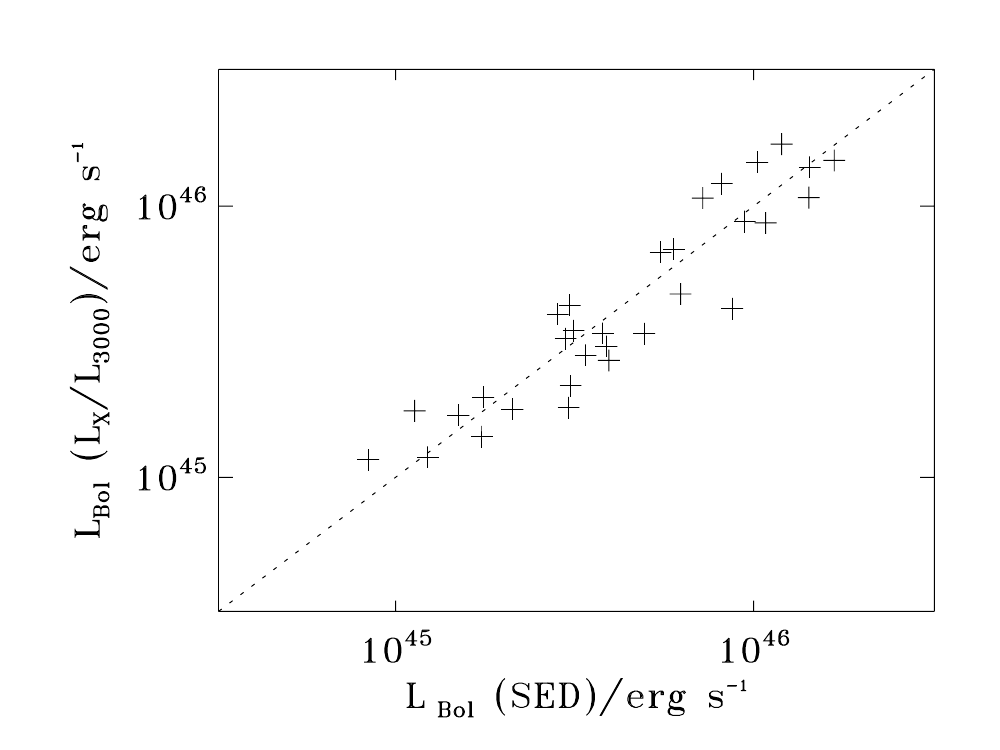}
\caption{Left - $\kappa_{2-10}$ vs. \lx/$L_{3000}$ for \nkbolsample\ sources in COSMOS with \lx, $L_{3000}$ and \lbol\ measured with SED fitting from L12. The line plotted through the data shows the results of a linear regression analysis. Right - Comparison of \lbol\ calculated from SED fitting in L12 to \lbol\ calculated using \lx\ and $\kappa_{\rm2-10}$ derived from the relationship between $\kappa_{\rm2-10}$ and \lx/$L_{3000}$. The line shows the one to one relationship.}
\label{aoxkbol}
\end{figure*}

The primary goal of this study is to investigate the relationship between $\Gamma$ and \lamedd, which requires a measurement of the bolometric luminosity. For \nlbolsed\ COSMOS sources, we take bolometric luminosities presented in L12, which are derived from spectral energy distribution (SED) fitting, from observed 160 \mic\ to hard X-rays. Bolometric luminosities for type 1 AGN are calculated in the rest frame 1 \mic\ to 200 keV. These data do not exist for our E-CDF-S sources however. For these sources we use a bolometric correction to the 2-10 keV X-ray luminosity ($\kappa_{2-10}$) to estimate \lbol. \cite{lusso10} showed that this can be done reliably using $\kappa_{2-10}$ derived from the X-ray to optical spectral index, \aox. \aox\ is normally calculated using the monochromatic luminosities at 2500 \AA\ and 2 keV, however, we utilise the 3000 \AA\ luminosity we already have at hand from the line measurements, measured from the optical spectra, and the 2-10 keV luminosity measured in the X-ray spectra and calibrate a relationship between $\kappa_{\rm2-10}$ and \lx/$L_{3000}$, using sources with known \lbol\ in COSMOS. We utilise \nkbolsample\ sources with $L_{3000}$ measured from FMOS and where \lx\ and \lbol\ are available for this analysis. Figure \ref{aoxkbol} shows that $\kappa_{\rm2-10}$ and \lx/$L_{3000}$ are indeed tightly correlated. We perform a linear regression analysis on these data, finding that log$_{10}\kappa_{2-10}=-0.70$log$_{10}$(\lx/$L_{3000}$)+0.84. We then calculate \lbol\ for the E-CDF-S  sources using this relationship. Figure \ref{aoxkbol} also shows the comparison between \lbol\ calculated in this manner and \lbol\ from L12 with good agreement between the two. For our sample of X-ray bright AGN to which we restrict our analysis with $\Gamma$, \neddsedxdata\ have \lbol\ from L12 and \neddaoxxdata\ have \lbol\ calculated using $\kappa_{\rm2-10}$.

\subsection{Radio properties of the sample}
\label{radio}

\begin{figure}
\includegraphics[width=90mm]{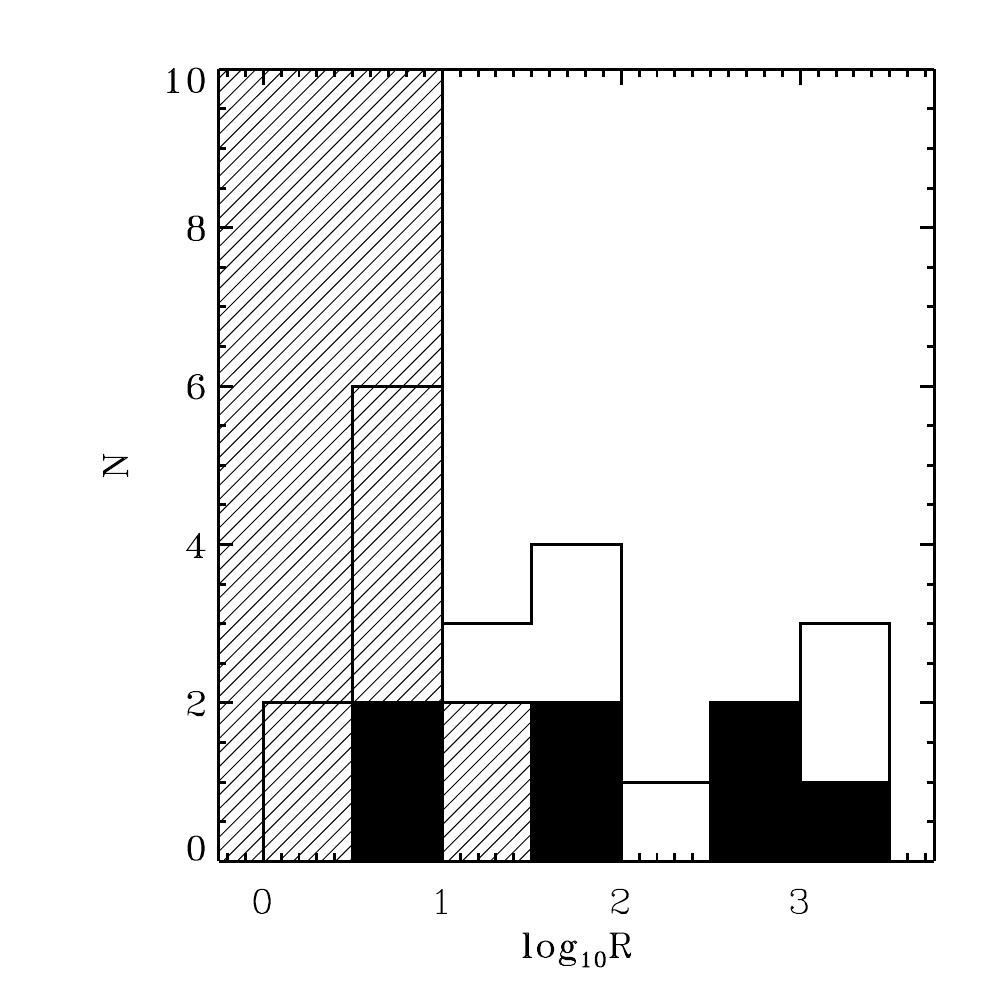}
\caption{Distribution of radio loudness parameter, R, for sources with radio detections (empty histograms), where the solid histogram is for E-CDF-S sources. The hatched histogram shows upper limits on the sources with no radio detection, assuming $10\umu$Jy sensitivity.}
\label{radio2}
\end{figure}

As the subject of this study is to probe the X-ray emission from the corona, we must account for other sources of X-ray emission intrinsic to the AGN. Radio loud AGN constitute $\sim$10\% of the AGN population, the fraction of which may vary with luminosity and redshift \citep[e.g.][]{jiang07} and are known to exhibit X-ray emission attributable to synchrotron or synchro-Compton emission from a jet \citep{zamorani81}. As such we must exclude radio loud AGN from our study. We compile radio data on our sample, specifically to exclude radio loud sources from our analysis, but also to investigate the radio properties. Deep ($\sim10\umu$Jy rms) VLA 1.4 GHz observations of both COSMOS and the E-CDF-S exist \citep{schinnerer07,miller13}. In COSMOS, we match the \xmm\ sources to the radio sources, and in E-CDF-S, we take the matches from \cite{bonzini12}. For \radiodet\ sources with a radio detection, we calculate the radio loudness parameter \citep{kellerman89}, R, which is defined as $f_{\rm 5 GHz}/f_{4400}$. We extrapolate the observed 1.4 GHz flux density to rest-frame 5 GHz using a power-law slope of -0.8. For the 4400\AA\ flux density, we extrapolate the observed R band flux density to rest-frame 4400\AA\ using a power-law slope of -0.5. R band magnitudes were taken from \cite{brusa10} and \cite{lehmer05} for COSMOS and E-CDF-S respectively. For radio undetected sources, we calculate an upper limit on R assuming a $10\umu$Jy sensitivity in both fields. Figure \ref{radio2} shows the distribution of R in the sample. We find six radio loud (R$>$100) sources in our sample, which we exclude from our analysis with $\Gamma$, leaving us with a sample of \neddxrqdata\ radio quiet AGN. Using upper limits, we can rule out radio loudness in the undetected sources. 

\section{What influences the physical conditions of the corona?}

It is understood that the characteristic shape of the X-ray spectrum produced in the corona, parametrised by $\Gamma$, depends on the electron temperature, T$_{e}$, and the optical depth to electron scattering, $\tau_{es}$, but it is unknown how these conditions arise, and how they relate to the accretion flow. We explore the dependence of $\Gamma$ on the various accretion parameters in an attempt to understand what brings about the physical conditions of the corona. The parameters we explore in this work are $L_{\rm UV}$, \lx, FWHM, \mbh\ and \lamedd.

\subsection{\lx\ and redshift}
\label{lxred}

\begin{figure}
\includegraphics[width=90mm]{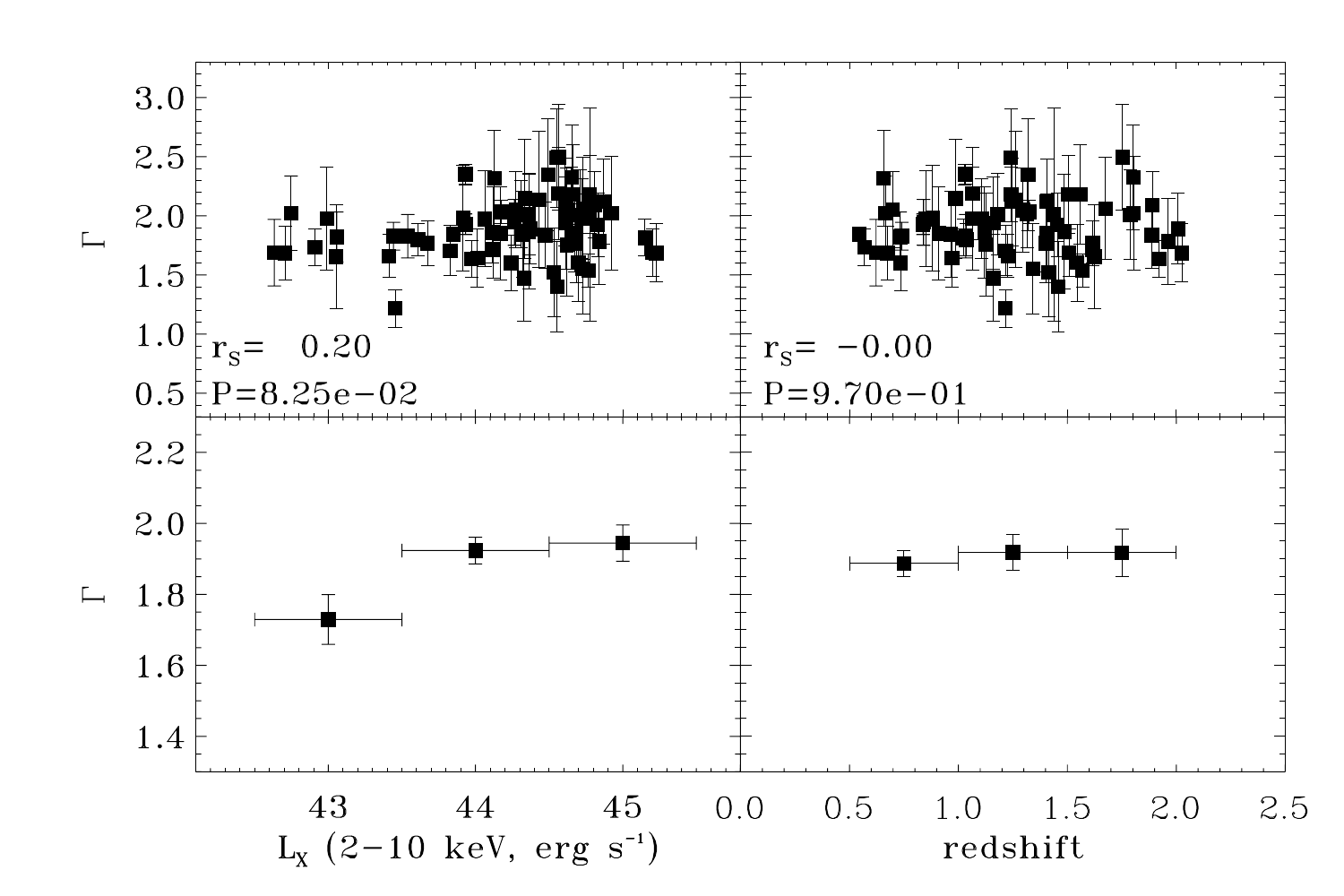}
\caption{Plot of $\Gamma$ vs \lx\ and redshift, where top panels show individual measurements and bottom panels show binned averages. Spearman rank correlation coefficients and p-values are displayed, showing there are no significant correlations with either \lx\ or redshift, however, the lowest luminosity bin shows systematically lower $\Gamma$ values than the higher luminosity bins.}
\label{gamma0}
\end{figure}

In order to investigate the relationships between $\Gamma$ and FWHM, \mbh\ and \lamedd, we first check for any dependencies on \lx\ and redshift in our sample in order to rule out degeneracies, as correlations with \lx\ and redshift have been reported previously \citep{dai04,saez08}. Figure \ref{gamma0} plots these relationships. A Spearman rank correlation analysis shows that there are no significant correlations present between $\Gamma$ and \lx\  or $\Gamma$ and redshift within this sample. We do note however that at the lowest X-ray luminosities (\lx$<3\times10^{43}$\ergs) that $\Gamma$ is systematically lower than at higher luminosities. While these are only 7 sources, we consider what affect if any they have on our results in a later section.

\subsection{$L_{\rm UV}$, FWHM, \mbh\ and \lamedd}

\begin{figure*}
\includegraphics[width=180mm]{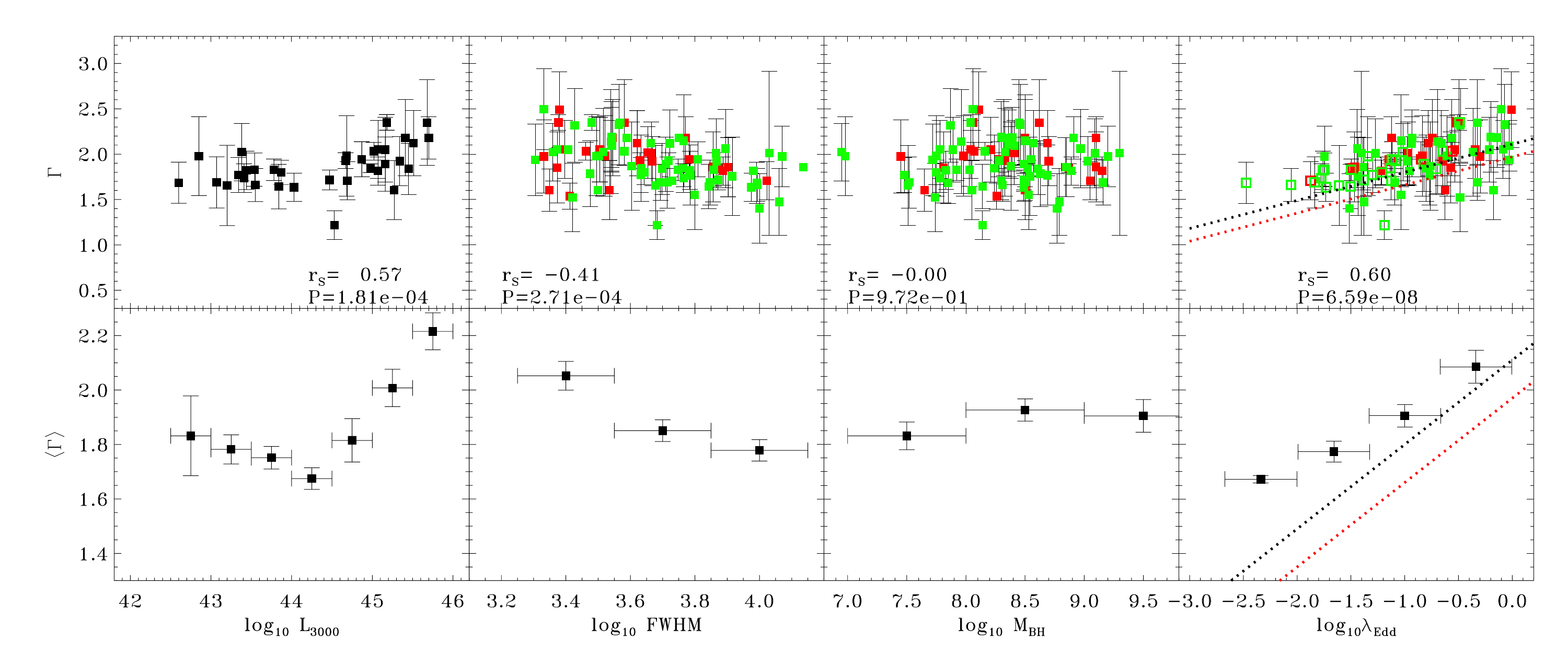}
\caption{Plots of $\Gamma$ versus UV luminosity ($\lambda$=3000\AA), FWHM (km s$^{-1}$), black hole mass ($M_{\odot}$) and \lamedd. Top panels show individual measurements, where red data points are measurements from the \ha\ line, and green from the \mgii\ line. For $\lambda_{\rm Edd}$, filed squares are data derived using \lbol\ from SED fitting, whereas empty squares are data derived using \lbol\ from \lx/$L_{3000}$. Bottom panels show binned averages. The Spearman-rank correlation coefficient, \rs, and the p-value are shown for each relationship. The black and red dotted lines are previous correlations reported on $\Gamma$-\lamedd\ from S08 and R09 respectively.}
\label{gammaxa}
\end{figure*}

We next investigate the dependence on the observed quantities, the 3000 \AA\ UV luminosity and FWHM of the lines and those derived quantities, black hole mass and \lamedd. The UV luminosity traces the accretion disc luminosity, and is thus related to the mass accretion rate, $\dot{m}$ ($L_{\rm acc}=\eta\dot{m}c^{2}$), whereas FWHM traces the gravitational potential. Furthermore, $L_{3000}$ and FWHM are the ingredients in the black hole mass calculation, which in turn is used in the determination of \lamedd. In Fig. \ref{gammaxa}, we show how $\Gamma$ depends on these four quantities.  

Here we are using a combined sample of \mgii\ and \ha\ measurements for FWHM, \mbh\ and \lamedd. In the figure we distinguish between the two line measurements with different colouring. When considering the combined sample, if a source has a measurement from both lines, we use the \ha\ measurement over the \mgii\ measurement. For \lamedd, we also differentiate between cases where \lbol\ has been determined using SED fitting (filled squares), or where it has been determined from \lx/$L_{3000}$ (open squares). 

A strong correlation with $L_{3000}$ is seen, which breaks at low luminosities ($L_{3000}\sim10^{44}$\ergs) and a strong anti-correlation with FWHM can be seen. These then cancel out here to give no dependence of $\Gamma$ on \mbh. A strong correlation is then seen with \lamedd. The Spearman rank correlation test shows that there is a significant correlation between $\Gamma$ and $L_{3000}$ ($r_{\rm S}=0.57$ and $p=1.81\times10^{-4}$, where $r_{\rm S}$ is the Spearman rank correlation coefficient and p is the probability of obtaining the absolute value of $r_{\rm S}$ at least as high as observed, under the assumption of the null hypothesis of zero correlation.), a significant anti-correlation ($p=$\pnhgamfwhm) between $\Gamma$ and the FWHM ($r_{\rm S}=$\rsgamfwhm) and a highly significant correlation ($p=$\pnhgamlam) between $\Gamma$ and \lamedd\ ($r_{\rm S}=$\rsgamlam). 

Despite the several ingredients used to derive \lamedd, this is by far the strongest correlation seen, stronger than with the observed quantities $L_{3000}$ and FWHM. This further confirms \lamedd\ as the primary parameter influencing the physical conditions of the corona responsible for the shape of the X-ray spectrum, being the electron temperature and electron scattering optical depth. Furthermore, as the relationship with $L_{3000}$, which is linked to the mass accretion rate, shows a break at low luminosities, which is not evident in the relationship with \lamedd, which is related to the mass accretion rate scaled by the black hole mass, this implies that the Eddington rate is more important than mass accretion rate.

Also shown on these plots are the $\Gamma-$\lamedd\ correlations previously reported by S08 and R09. Our results are consistent with S08 in the two highest bins of \lamedd, however our results diverge at lower values of \lamedd, with a flatter $\Gamma-$\lamedd\ relationship. Our results are systematically higher than those reported in R09. We explore these differences further in section \ref{compgamlamcor}.

\begin{figure}
\includegraphics[width=90mm]{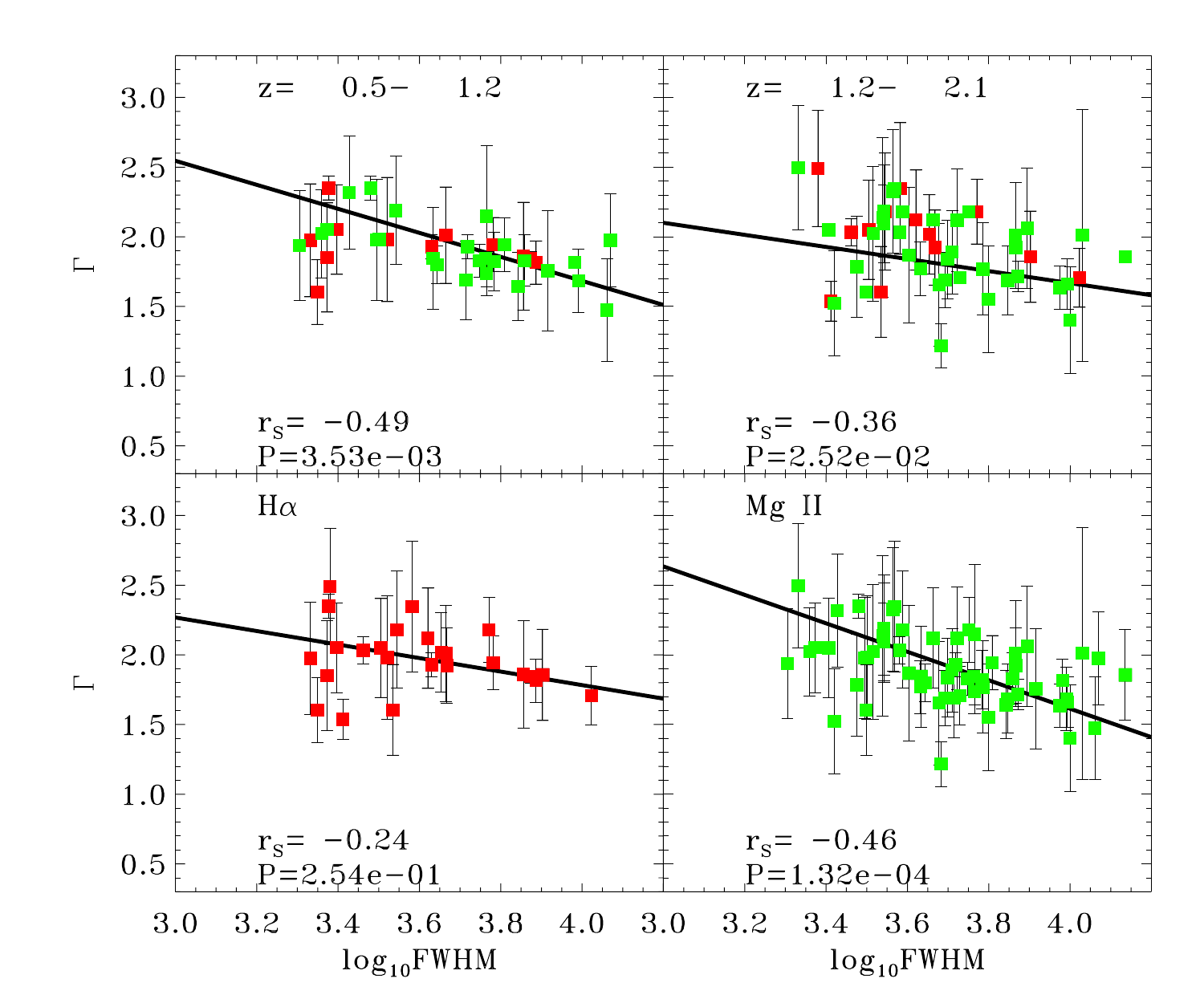}
\caption{Plots of $\Gamma$ vs the FWHM of the emission lines in four subsamples: two redshift bins, \redrangea\ and \redrangeb\ (top panels) and for the measurements from \ha\ and \mgii\ separately (bottom panels).}
\label{gammafw}
\end{figure}

For the two relationships between $\Gamma$ and FWHM and $\Gamma$ and \lamedd\ where emission line measurements were used, we further the investigation in different subsamples:  in two different redshift bins (\redrangea\ and \redrangeb), effectively splitting the sample in half, and considering measurements made by \ha\ and \mgii\ separately. Table \ref{fwhm_table} presents the results of Spearman rank and linear regression analysis of the $\Gamma$ vs. FWHM relationship for these different subsamples. The correlation is most significant in the full redshift range when using only the \mgii\ measurements ($p=$\pnhgamfwhmmg), however the correlation is not significant ($p>0.1$) when considering only \ha\ measurements, though this may be due to small number statistics. For the whole sample, we find that 
\begin{equation}
\Gamma=(0.32\pm0.05){\rm log}_{10}\lambda_{\rm Edd}+(2.27\pm0.06)
\end{equation}
Taking the two line measurements separately gives \reggamfwhmha\ from \ha\ and \reggamfwhmmg\ from \mgii. The slopes are significantly different, at  $>$3-$\sigma$. Fig. \ref{gammafw} shows the sample in the two redshift bins and for the two line measurements separately, along with the best fit lines.  

\begin{table*}
\centering
\caption{Table of Spearman rank correlation and linear regression analysis for $\Gamma$ vs.log$_{10}$FWHM for sources with a measurement from \mgii\ or \ha\ and with greater than 250 source counts in the X-ray spectrum. In the combined sample of \mgii\ and \ha\ measurements, if both exist for one source, the \ha\ measurement is used over the \mgii\ measurement. Column (1) is the redshift range used; (2) is the emission lines used for the FWHM; (3) is the total number of sources within each subsample; (4) is the Spearman rank coefficient; (5) is the null hypothesis probability; (6) is the gradient coefficient in the linear regression analysis, where $\Gamma=$m log$_{10}$FWHM+c; and (7) is the constant in this relationship.}
\label{fwhm_table}
\begin{center}
\begin{tabular}{ l l l l l l l }
\hline
redshift range & lines used & number in subsample & $R_{\rm S}$ & $p$ & m & c \\
(1) & (2) & (3) & (4) & (5) & (6) & (7) \\
\hline
   0.5-   2.1 & H$\alpha$ \& MgII & 73 &  -0.41 & 2.71$\times10^{-4}$ &  -0.69$\pm$  0.11 &   4.44$\pm$  0.42\\
& H$\alpha$ & 25 &  -0.24 & 2.54$\times10^{-1}$ &  -0.49$\pm$  0.14 &   3.72$\pm$  0.53\\
& MgII & 65 &  -0.46 & 1.32$\times10^{-4}$ &  -1.02$\pm$  0.15 &   5.70$\pm$  0.54\\
   0.5-   1.2 & H$\alpha$ \& MgII & 34 &  -0.49 & 3.53$\times10^{-3}$ &  -0.86$\pm$  0.14 &   5.13$\pm$  0.52\\
& H$\alpha$ & 12 &  -0.25 & 4.30$\times10^{-1}$ &  -0.71$\pm$  0.17 &   4.58$\pm$  0.64\\
& MgII & 29 &  -0.63 & 2.88$\times10^{-4}$ &  -1.23$\pm$  0.18 &   6.48$\pm$  0.68\\
   1.2-   2.1 & H$\alpha$ \& MgII & 39 &  -0.36 & 2.52$\times10^{-2}$ &  -0.43$\pm$  0.19 &   3.40$\pm$  0.71\\
& H$\alpha$ & 13 &  -0.20 & 5.17$\times10^{-1}$ &  -0.01$\pm$  0.33 &   1.97$\pm$  1.17\\
& MgII & 36 &  -0.35 & 3.68$\times10^{-2}$ &  -0.61$\pm$  0.24 &   4.09$\pm$  0.91\\
\hline
\end{tabular}
\end{center}
\end{table*}

\begin{figure}
\includegraphics[width=90mm]{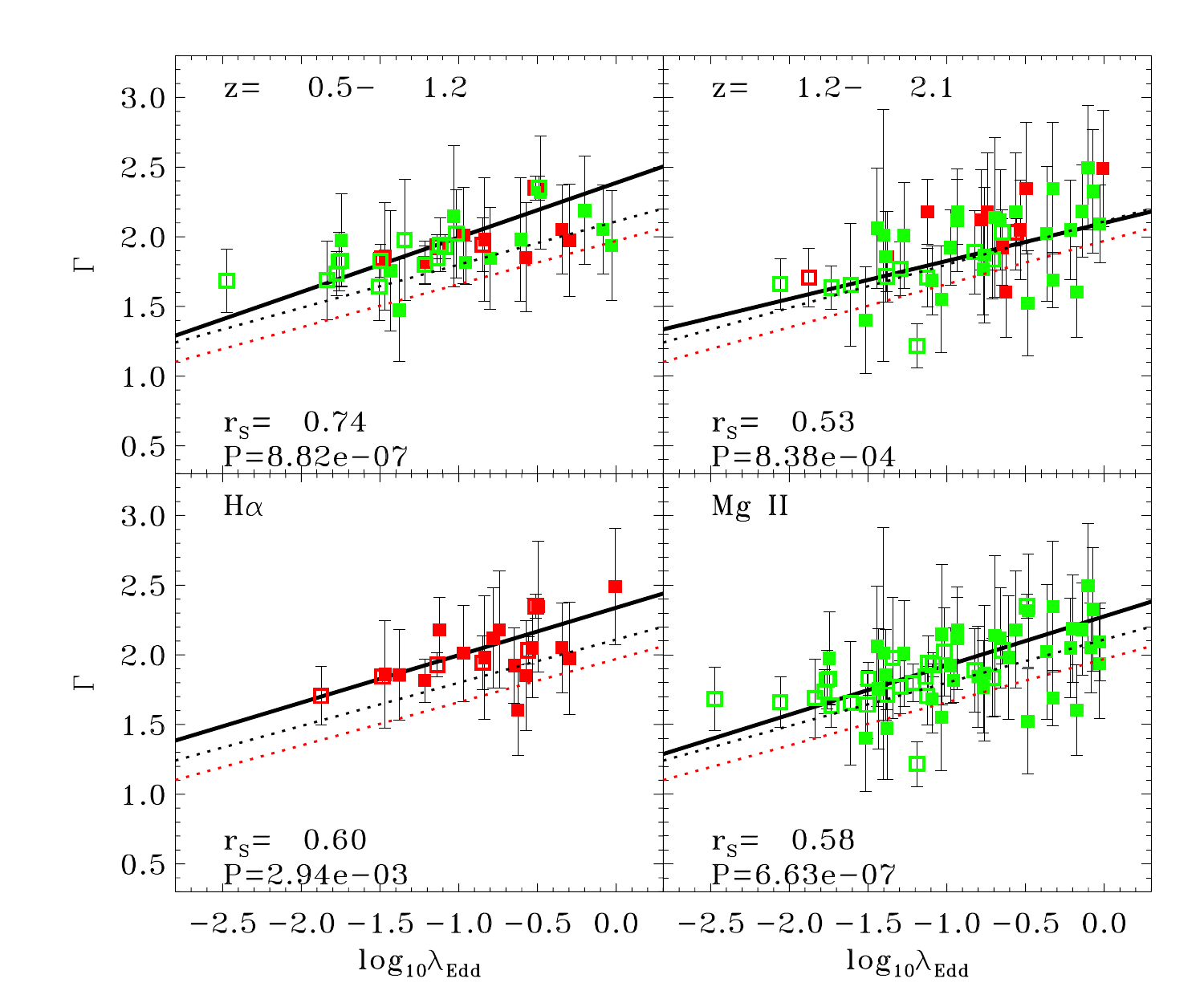}
\caption{Plots of $\Gamma$ vs  \lamedd\ for four subsamples: two redshift bins, \redrangea\ and \redrangeb\ (top panels) where red data points are measurements from the \ha\ line, and green from the \mgii\ line. Filled symbols are data derived using \lbol\ from SED fitting, whereas empty symbols are data derived using \lbol\ from \lx/$L_{3000}$. The bottom panels show the measurements from \ha\ and \mgii\ separately in the full redshift range. The black and red dotted lines are previous correlations reported on $\Gamma$-\lamedd\ from S08 and R09.}
\label{gammalam}
\end{figure}

We also carry out Spearman rank correlation analysis and linear regression analysis on the $\Gamma$ vs. \lamedd\ relationship for the different subsamples. The results are presented in Table \ref{lam_table}. The correlation is most significant in the full redshift range when using the two line measurements together ($p=$\pnhgamlam). The correlation is significant ($p<0.1$) in all redshift bins, however, not for measurements made with \ha\ in the highest redshift bin. This is again likely due to small number statistics. For the whole sample we find that 
\begin{equation}
\Gamma=( -0.69\pm  0.11){\rm log}_{10}{\rm(FWHM/km s}^{-1})+(  4.44\pm  0.42)
\end{equation}
The slope of the correlations for \ha\ and \mgii\ agree very well with each other with \reggamlamha\ for \ha\ and \reggamlammg\ for \mgii\ and for all redshift bins. This supports the claim by \cite{matsuoka13} that the two line measurements give consistent black hole mass estimates. We plot $\Gamma$ versus \lamedd\ in the two redshift bins and for the two lines separately in Figure \ref{gammalam}, along with the best fit lines.

\begin{table*}
\centering
\caption{Table of Spearman rank correlation and linear regression analysis for $\Gamma$ vs. log$_{10}$\lamedd, for sources with \mbh\ from \mgii\ or \ha, an estimate of the bolometric luminosity and those with greater than 250 source counts in the X-ray spectrum. In the combined sample of \mgii\ and \ha\ measurements, if both exist for one source, the \ha\ measurement is used over the \mgii\ measurement. Column (1) is the redshift range used; (2) is the emission lines used for the estimation of the black hole mass; (3) is the total number of sources within each subsample; (4) is the Spearman rank coefficient; (5) is the null hypothesis probability; (6) is the gradient coefficient in the linear regression analysis, where $\Gamma$=m log$_{10}$\lamedd+c; and (7) is the constant in this relationship.}
\label{lam_table}
\begin{center}
\begin{tabular}{ l l l l l l l }
\hline
redshift range & lines used & number in subsample & $R_{\rm S}$ & $p$ & m & c \\
(1) & (2) & (3) & (4) & (5) & (6) & (7) \\
\hline
   0.5-   2.1 & H$\alpha$ \& MgII & 69 &   0.60 & 6.59$\times10^{-8}$ &   0.32$\pm$  0.05 &   2.27$\pm$  0.06\\
& H$\alpha$ & 22 &   0.60 & 2.94$\times10^{-3}$ &   0.34$\pm$  0.07 &   2.34$\pm$  0.09\\
& MgII & 64 &   0.58 & 6.63$\times10^{-7}$ &   0.35$\pm$  0.05 &   2.28$\pm$  0.06\\
   0.5-   1.2 & H$\alpha$ \& MgII & 33 &   0.74 & 8.82$\times10^{-7}$ &   0.39$\pm$  0.06 &   2.39$\pm$  0.07\\
& H$\alpha$ & 11 &   0.67 & 2.33$\times10^{-2}$ &   0.43$\pm$  0.09 &   2.46$\pm$  0.11\\
& MgII & 29 &   0.75 & 2.82$\times10^{-6}$ &   0.43$\pm$  0.07 &   2.39$\pm$  0.07\\
   1.2-   2.1 & H$\alpha$ \& MgII & 36 &   0.53 & 8.38$\times10^{-4}$ &   0.27$\pm$  0.08 &   2.10$\pm$  0.09\\
& H$\alpha$ & 11 &   0.50 & 1.17$\times10^{-1}$ &   0.23$\pm$  0.15 &   2.18$\pm$  0.14\\
& MgII & 35 &   0.52 & 1.22$\times10^{-3}$ &   0.27$\pm$  0.08 &   2.08$\pm$  0.09\\
\hline
\end{tabular}
\end{center}
\end{table*}

\subsection{Radio loud sources}

\begin{figure}
\includegraphics[width=90mm]{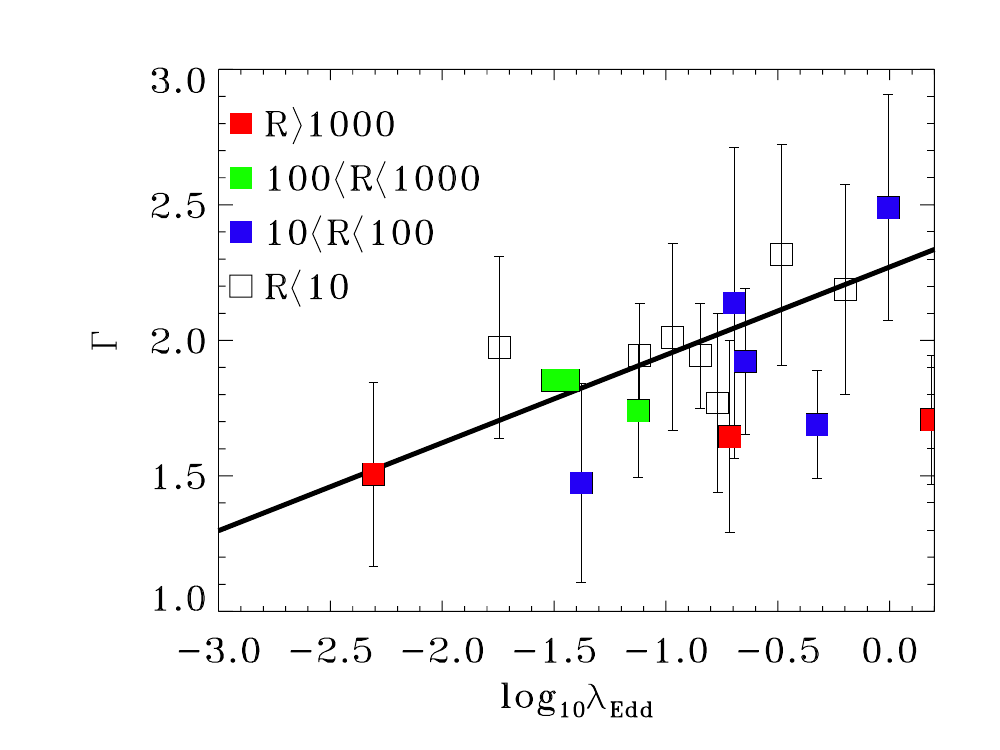}
\caption{$\Gamma$ vs. \lamedd\ for sources with radio detections only, where colour corresponds to radio loudness, parametrised by R. The solid line is the best fit line to the full sample found in the previous section, when excluding radio loud sources. We find that radio loud sources are generally consistent with this trend, with the exception of sources with high Eddington ratios (\lamedd$>0.3$).}
\label{gamradio}
\end{figure}

In section \ref{radio}, we investigated the radio properties of the sample, finding six sources which are radio loud (R$>$100). We excluded these sources from our sample as radio loud AGN have significant X-ray emission from their jets, and thus contaminate the coronal emission which interests us. However, we briefly explore the properties of the radio loud sources here. Fig \ref{gamradio} plots the $\Gamma$-\lamedd\ relationship found in the previous section, here with only sources with radio detections. We colour code the data points by radio loudness. We find that radio loud sources are generally consistent with this trend, with the exception of two sources with high Eddington ratios (\lamedd$>0.3$).


\section{discussion}

\subsection{Accounting for degeneracies and biases}

\subsubsection{Low \lx\ sources}

We noted in section \ref{lxred} that at low X-ray luminosities (\lx$<3\times10^{43}$ \ergs), $\Gamma$ is systematically lower. While there are only 7 sources at these low luminosities, and a Spearman rank correlation test tells us there is a lack of a significant correlation between $\Gamma$ and \lx\ in our sample, we check the effects on our results when excluding these sources at low \lx. When doing this, we find that a significant correlation between $\Gamma$ and \lamedd\ persists, with $p=7.7\times10^{-6}$. Linear regression analysis of this subsample gives $\Gamma=( 0.35\pm0.05)$log$_{10}$\lamedd$+(2.30\pm0.06)$. We plot these data with the best fit trend line in Fig. \ref{gammalam2}.

\subsubsection{The soft X-ray excess and FWHM degeneracy}

When it was reported by \cite{boller96} and \cite{laor97} that the soft X-ray (0.2-2 keV) power-law index correlated well with the FWHM of the \hb\ line, also interpreted as a dependence on accretion rate, it was thought that this may have been related to the soft X-ray excess seen in unobscured AGN \citep[e.g.][]{arnaud85,turner89}. This feature is strong in narrow line Seyfert 1s \citep{boller96}, which have low FWHMs \citep{osterbrock85}, and as such a strong soft excess could cause a steepening of the X-ray spectrum leading to higher values of $\Gamma$ for low FWHM sources. \cite{brandt97} later reported a correlation between $\Gamma$ in the 2-10 keV band and the FWHM of the \hb\ line, where this band is less affected by the soft excess, leading to the conclusion that both the soft excess and the power-law are affected by the accretion rate. We attempt to further rule out the effect of the soft excess on the power-law by considering the 4-10 keV rest-frame band where the X-ray spectrum should be completely independent of the soft excess. We rerun our analysis described in section \ref{xspec} with this restriction, however, in doing this we decrease the number of spectral counts. We therefore lower the count cut that we make using the 2-10 keV spectrum of 250 counts to 100 counts in the 4-10 keV band. Despite this we still find a significant correlation between $\Gamma$ and \lamedd\ with a p-value of 0.002. This confirms that the soft excess does not substantially contribute to the $\Gamma$-\lamedd\ relationship.

We next consider the relationship between FWHM and \lamedd\ \citep{boroson92}, and the degeneracy it may introduce in our work. As our aim here is to rule out degeneracies where possible, we investigate the effect of making a cut in FWHM. We note that the correlation of $\Gamma$ vs. FWHM appears to be driven by sources with FWHM$<4000$ km s$^{-1}$, and thus we investigate the $\Gamma$-\lamedd\ relationship for sources with FWHM$>4000$ km s$^{-1}$ in our sample. In doing so, however, we still find a significant correlation between $\Gamma$ and \lamedd\ with a p-value of 0.005, which effectively rules out a bias from low FWHM sources, and rules out degeneracy with FWHM. A linear regression analysis finds that $\Gamma=(0.12\pm0.07)$log$_{10}$\lamedd$+(1.98\pm0.10)$. We plot the subsample with the best fit trend line in Fig. \ref{gammalam2}.

\subsubsection{Contribution from reflection}

It is important to consider what effects reflection of X-rays, from either the accretion disc or the circumnuclear torus may have on our results. In our X-ray spectral fitting we use a simple power-law to characterise the spectrum, however in reality reflection features are present.  As the geometry of the torus is expected to change with X-ray luminosity and redshift \citep[e.g.][]{lawrence91,ueda03,hasinger08,brightman12b}, this component should not be neglected. In order to account for this, we utilise the X-ray spectral torus models of \cite{brightman11}, which describe the X-ray spectra of AGN surrounded by a torus of spherical geometry. In order to explain the decrease in the AGN obscured fraction with increasing X-ray luminosity, and the increase with redshift, new work from Ueda, et al (in preparation) have calculated the dependence of the torus opening angle on these parameters. Essentially the torus opening angle increases with increasing X-ray luminosity and decreases with increasing redshift. The effect of this on the observed $\Gamma$ is for $\Gamma$ to decrease towards smaller opening angles, and hence greater covering factors, due to greater reflection from the torus which has a flat spectrum. We use this prescription when fitting our spectra with this torus model, where the viewing angle is set such that the source is unobscured, the \nh\ through the torus is set to $10^{24}$ \cmsq\ and the opening angle depends on the X-ray luminosity and redshift. We follow the same spectral fitting technique as described in section \ref{xspec}, and study the results. We find that the correlation between $\Gamma$ and \lamedd\ remains highly significant with a p-value of $2.5\times10^{-5}$. We note that the observed $\Gamma$ produced by this torus model changes by a maximum of 0.06 in the 2-10 keV range for the extremes of the parameter space, while we observe changes of greater than 0.2. It is therefore unlikely that reflection affects our results. S08 also investigated the effects of reflection in their analysis, finding only two sources where a reflection component was significantly detected. This low detection rate is expected due to the high luminosity nature of the sources at high redshifts.

\subsubsection{X-ray variability}

We briefly check if X-ray variability has an effect on our results, specifically if the outliers in our relationships may be explained by this. Lanzuisi, et al (in preparation) have conducted an investigation into AGN variability for XMM-COSMOS sources. They find that 6 of the sources in our sample are variable in X-rays through the detection of excess variance. We find however, that these sources lie consistently on the best fit relations found here, leading us to the conclusion that variability is not likely to affect our results. \cite{papadakis09} directly investigated the relationship between $\Gamma$ and AGN variability, finding that $\Gamma$ correlates with the characteristic frequency in the power spectrum when normalised by the black hole mass. They then used the result of \cite{mchardy06}, which showed that this normalised characteristic frequency is correlated to the accretion rate, to also show in an independent manner to what we have shown here, that $\Gamma$ is correlated with accretion rate. Their result held true even when using the mean spectral slope of their data. This result is relevant here, as we have used time averaged spectra, which \cite{papadakis09} have shown produces consistent results to time resolved spectroscopy.

\subsubsection{Chandra/XMM-Newton cross-calibration}

\begin{figure}
\includegraphics[width=90mm]{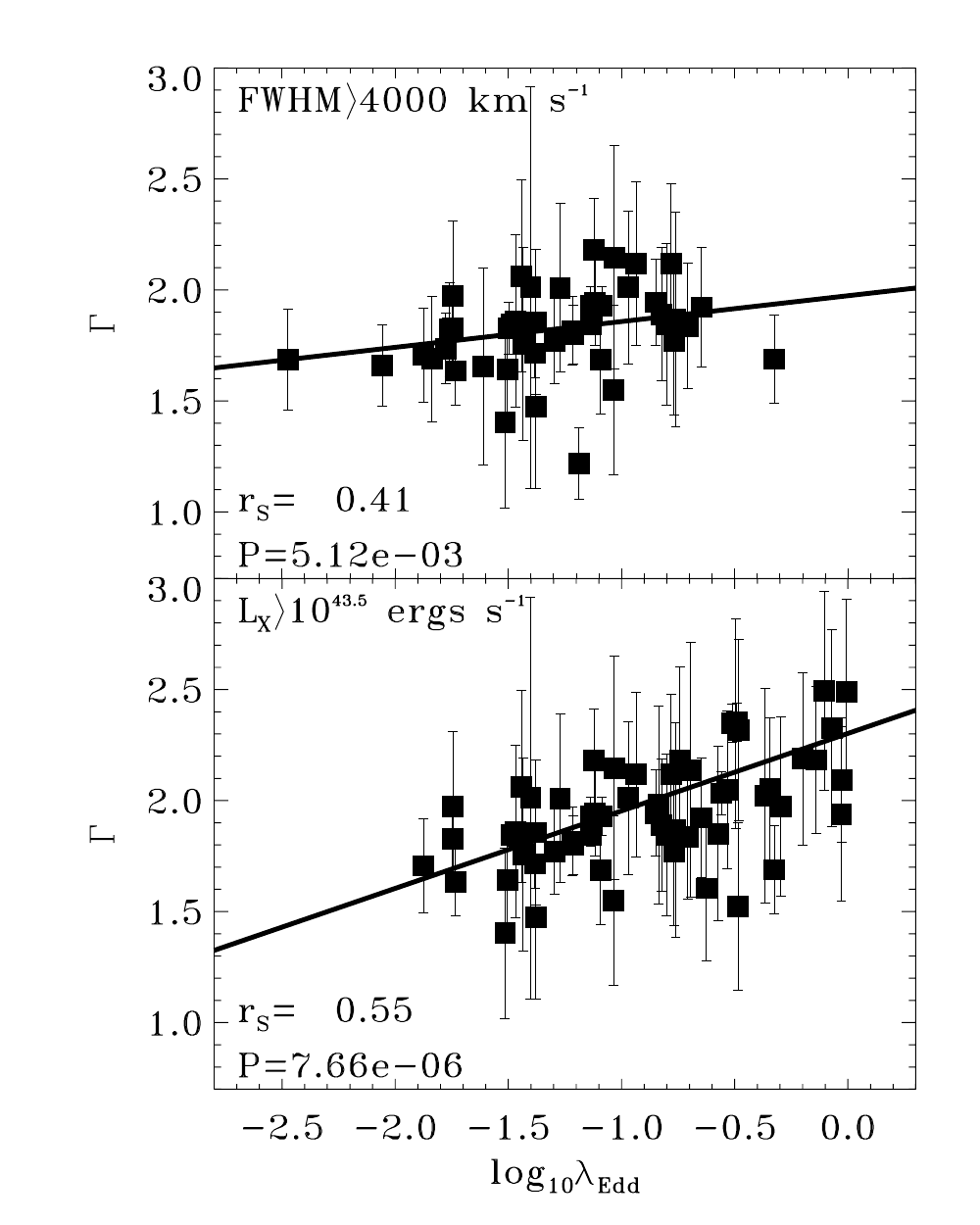}
\includegraphics[width=90mm]{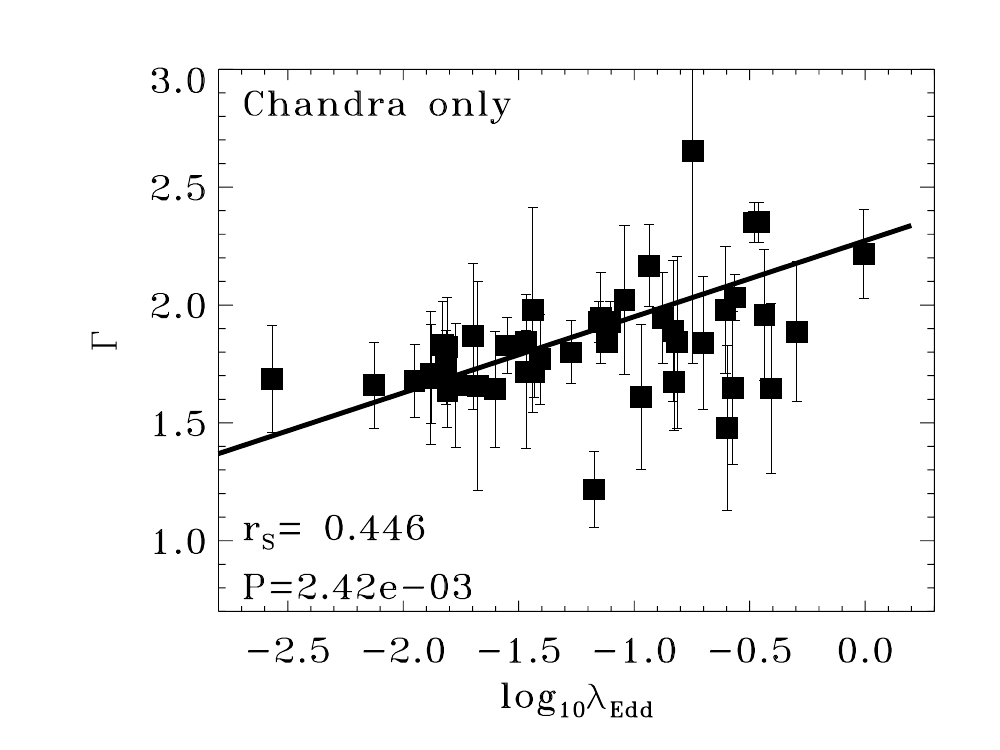}
\caption{Plots of $\Gamma$ vs  \lamedd\ when excluding sources with FWHM$<$4000 km s$^{-1}$ (top) and sources with \lx$<10^{43.5}$ \ergs\ (middle) in order to break any degeneracy with these parameters. The lowest panel shows  $\Gamma$ vs  \lamedd\ when using \chandra\ data in COSMOS instead of \xmm\ data in order to investigate the effect of the cross-normalisation between these telescopes.}
\label{gammalam2}
\end{figure}

We use both \chandra\ and \xmm\ data in our analysis, with \xmm\ data in COSMOS and \chandra\ data in the E-CDF-S . However, it has been found that a systematic difference between measurements made by the two observatories of the same source exists (L13). Most relevant to our work is a systematic difference of up to 20\% in $\Gamma$, which may seriously bias our results. We investigate this issue by performing our analysis using the \chandra\ data available in COSMOS. We extract the \chandra\ spectra as described in section \ref{cdfs} for the E-CDF-S  sources and analyse the data in the same way. While the sample size is reduced to 44 sources with $\Gamma$ and \lamedd\ due to the smaller coverage of the \chandra\ observations in COSMOS and the lower number of source counts per spectrum, our main result is maintained. For $\Gamma$ vs. \lamedd, the Spearman rank correlation analysis reveals a significant correlation with $p=2.42\times10^{-3}$ and linear regression analysis finds that $\Gamma=(  0.32\pm  0.05)$log$_{10}$\lamedd$+(  2.27\pm  0.06)$, which is in very good agreement with the joint \chandra/\xmm\ analysis. These data are plotted with the best fitting trend line in Fig. \ref{gammalam2}.

\subsection{Comparison with previous studies}

\subsubsection{X-ray spectral analysis in COSMOS}

Recent work by L13 has presented a spectral analysis of bright \chandra-COSMOS sources, the aim of which was to determine the intrinsic absorption in the spectrum, as well as $\Gamma$. Their work also presents analysis of \xmm\ spectra of the \chandra\ counterparts. Their analysis differs from our own as they utilise the fuller 0.5-7 keV band pass and sources with greater than 70 net counts, whereas we limit ourselves to analysis in the rest-frame 2-10 keV band and spectra with at least 250 counts. We investigate the differences in these methods by comparing the results for 129 common sources, in particular with respect to absorption. As we restrict ourselves to rest-frame energies greater than 2 keV, we are not sensitive to \nh$\lesssim10^{22}$ \cmsq, whereas L13 utilise a fuller band pass and as such are thus sensitive to lower levels of absorption. While we only detect absorption in one source from our COSMOS sample, this is not a common source with L13. From the common sources, they detect absorption in six sources (XMM-IDs 28, 30, 34, 38, 66 and 71), however the measured \nh\ is $<5\times10^{21}$ \cmsq\ in all of these, and thus will have no affect on our measurement of $\Gamma$. When comparing $\Gamma$ measurements between the two analyses, there is good agreement, with the difference being within our measurement error and there being no systematic offsets.

\subsubsection{$\Gamma$-\lamedd\ correlation}
\label{compgamlamcor}

In Figs. \ref{gammaxa} \& \ref{gammalam}, we have compared our results on the $\Gamma$-\lamedd\ correlation with two previous works by S08 and R09. We found that for the whole sample, our binned averages were consistent with the results of S08 in the highest \lamedd\ bins, while overall systematically higher than those of R09. At lower \lamedd\ values our $\Gamma$ measurements lie above these previous correlations.

The results from S08 were based on a sample of 35 moderate- to high-luminosity radio quiet AGN having high quality optical and X-ray spectra, based on individual observations, and including nearby sources and those up to z=3.2. The black hole masses were estimated from the \hb\ line. The results of R09 are based on 403 sources from the {\it SDSS/XMM-Newton} quasar survey of \cite{young09}, which span a wide range in X-ray luminosity ($10^{43}<$\lx$<10^{45.5}$ \ergs) and redshift (0.1$<z<4.5$), with black hole masses based on \hb, \mgii\ and C {\sc iv}.  The work presented here is based on a sample of \neddxrqdata\ sources with black hole estimates based on \ha\ and \mgii, and samples X-ray luminosities in the range of $10^{42.5}<$\lx$<10^{45.5}$ \ergs\ and the redshift range of $0.5<z<2.1$. 

Our analysis finds that \reggamlam. Meanwhile, S08 find that $\Gamma=(0.31\pm0.01)$log$_{10}$\lamedd$+(2.11\pm0.01)$ and R09 find that $\Gamma=(0.31\pm0.06)$log$_{10}$\lamedd$+(1.97\pm0.02)$. The slope of our $\Gamma$-\lamedd\ correlation is in excellent agreement with these two previous works. The constant in the relationship is higher than both studies by 3-$\sigma$. This may be due to differing methods of X-ray spectral fitting. Here we use Cash statistics in our analysis, whereas previous works have used $\chi^2$, the two being known to produce systematic differences in $\Gamma$ \citep{tozzi06}.

Furthermore, R09 reported significantly differing slopes for $\Gamma-$\lamedd(\hb) and $\Gamma-$\lamedd(\mgii), where the \hb\ slope was steeper. We also compared our slopes for \ha\ and \mgii, but found no such difference. We make a comparison to the results from S08 and R09 considering the different lines separately in Table \ref{lam_table2}.  For the \mgii\ line, only R09 have presented results and we find that our results are consistent with these within 2-$\sigma$. We present here for the first time results based on \ha\ alone, however, our \ha\ results are in very good agreement with the \hb\ results in S08, though the difference between these results and the \hb\ results in R09 is 3-$\sigma$.

The differing results between \hb\ and \mgii\ reported in R09 were attributed to uncertainties in the measurement of \mgii. However, \cite{matsuoka13}, which describes the measurements of \ha\ and \mgii\ used in this sample find very good agreements between these lines, which subsequently has lead to the good agreement we find between them here.

\begin{table}
\centering
\caption{A summary of the results of the best fit line to the $\Gamma$-\lamedd\ correlation from this work and those of S08 and R09, for line measurements made from \ha, \hb\ and \mgii. Column (1) is the sample used, column (2) is the lines used, column (3) is the slope of the relationship and column (4) is the intercept}
\label{lam_table2}
\begin{center}
\begin{tabular}{ l l l l}
\hline
Sample & Lines & m & c \\
(1) & (2) & (3) & (4) \\
\hline
This work & \ha\ \& \mgii\ & \reggamlamm\ & \reggamlamc\ \\
R09 & \hb, \mgii\ \& C {\sc iv} & $0.31\pm0.06$ &$1.97\pm0.02$ \\
This work & \mgii\ & \reggamlammgm\ & \reggamlammgc\ \\
R09 & \mgii\ & $0.24\pm0.05$ & $1.98\pm0.02$\\
This work & \ha\  & \reggamlamham\ & \reggamlammgc\ \\
S08 & \hb\ & $0.31\pm0.01$ & $2.11\pm0.01$ \\
R09 & \hb\ & $0.58\pm0.11$ & $1.99\pm0.04$ \\
\hline

\end{tabular}
\end{center}
\end{table}

\subsection{Wider implications of our results}

We have confirmed previous results that \lamedd, rather than $L_{\rm UV}$, \lx, FWHM or \mbh, is the parameter which most strongly correlates with the X-ray spectral index, $\Gamma$, and hence is responsible for driving the physical conditions of the corona responsible for the shape of the X-ray spectrum, being the electron temperature and electron scattering optical depth. This result enables models of accretion physics in AGN to be constrained. As this correlation is stronger than the $\Gamma$-$L_{3000}$ correlation, where $L_{3000}$ is related to the mass accretion rate, $\dot{m}$ and \lamedd\ is related to the mass accretion rate scaled by \mbh, then the coronal conditions must depend on both mass accretion rate and black hole mass, despite there being no correlation with black hole mass itself. A possible interpretation of the correlation between $\Gamma$ and \lamedd\ is that as \lamedd\ increases, the accretion disc becomes hotter, with enhanced emission. This enhanced emission cools the electron corona more effectively, leading to a lower election temperature, electron scattering optical depth, or both. $\Gamma$ thus increases as these quantities decrease.
  
As pointed out by previous authors on this subject, a statistically significant relationship between $\Gamma$ and \lamedd\ allows for an independent estimate of \lamedd\ for AGN from their X-ray spectra alone. While the dispersion in this relationship means that this is not viable for single sources, it could be used on large samples, for example those produced by {\it eROSITA}. {\it eROSITA} is due for launch in 2014 and will detect up to 3 million AGN with X-ray spectral coverage up to 10 keV. These results could be valuable in placing estimates on the accretion history of the universe using these new data.

\section{Conclusions}

We have presented an X-ray spectral analysis of broad-lined radio-quiet AGN in the extended {\it Chandra} Deep Field South and COSMOS surveys with black hole mass estimates. The results are as follows:
\begin{itemize}
\item We confirm a statistically significant correlation between the rest-frame 2-10 keV photon index, $\Gamma$, and the FWHM of the optical broad emission lines, found previously by \cite{brandt97}. A linear regression analysis reveals that \reggamfwhm. 
\item A statistically signifiant correlation between $\Gamma$ and \lamedd\ is also confirmed, as previously reported in S06, S08 and R09. The relationship between $\Gamma$ and \lamedd\ is highly significant with a chance probability of \pnhgamlam. A linear regression analysis reveals that \reggamlam, in very good agreement with S08 and R09. The correlation with \lamedd\ is the strongest of all parameters tested against $\Gamma$, indicating that it is the Eddington rate, which is related to the mass accretion rate scaled by the black hole mass, that drives the physical conditions of the corona responsible for the X-ray emission. We find no statistically significant relationship between $\Gamma$ and black hole mass by itself.
\item We present this analysis for the first time at high redshift with the \ha\ line, finding that the $\Gamma$-\lamedd\ correlation agrees very well for measurements with \ha\ and with \mgii. The \mgii\ correlation is consistent with the results of R09 within 3-$\sigma$ and the \ha\ correlation is consistent with the \hb\ results from S08.
\end{itemize}

\bibliographystyle{mn2e}
\bibliography{bibdesk}


\label{lastpage}
\end{document}